# Incorporating the effect of heterogeneous surface heating into a semi-empirical model of the surface energy balance closure

Luise Wanner[1], Marc Calaf[2], Matthias Mauder[1,3,4]

[1]Institute of Meteorology and Climate Research – Atmospheric Environmental Research, Karlsruhe Institute of Technology, Garmisch-Partenkirchen, 82467, Germany
[2]Department of Mechanical Engineering, University of Utah, Salt Lake City, 84112, United States of America
[3]Institute of Geography and Geoecology, Karlsruhe Institute of Technology, Karlsruhe, 76131, Germany
[4]Institute of Hydrology and Meteorology, Technische Universität Dresden, Tharandt, 01737 Germany

*Correspondence to*: Luise Wanner (luise.wanner@kit.edu)

**Abstract.** It was discovered several decades ago that eddy covariance measurements systematically underestimate sensible and latent heat fluxes, creating an imbalance in the surface energy budget. Since then, many studies have addressed this problem and proposed a variety of solutions to the problem, including improvements to instruments and correction methods applied during data postprocessing. However, none of these measures have led to the complete closure of the energy balance gap. The leading hypothesis is that not only surface-attached turbulent eddies but also sub-mesoscale atmospheric circulations contribute to the transport of energy in the atmospheric boundary layer, and the contribution from organized motions has been grossly neglected. The problem arises because the transport of energy through these secondary circulations cannot be captured by the standard eddy covariance method given the relatively short averaging periods of time (~30 minutes) used to compute statistics. There are various approaches to adjust the measured heat fluxes by attributing the missing energy to the sensible and latent heat flux in different proportions. However, few correction methods are based on the processes causing the energy balance gap. Several studies have shown that the magnitude of the energy balance gap depends on the atmospheric stability and the heterogeneity scale of the landscape around the measurement site. Based on this, the energy balance gap within the surface layer has already been modelled as a function of a nonlocal atmospheric stability parameter by performing a large-eddy simulation study with idealized homogeneous surfaces. We have further developed this approach by including thermal surface heterogeneity in addition to atmospheric stability in the parameterization. Specifically, we incorporated a thermal heterogeneity parameter that was shown to relate to the magnitude of the energy balance gap. For this purpose, we use a Large-Eddy Simulation dataset of 28 simulations with seven different atmospheric conditions and three heterogeneous surfaces with different heterogeneity scales as well as one homogeneous surface. The newly developed model captures very well the variability in the magnitude of the energy balance gap under different conditions. The model covers a wide range of both atmospheric stabilities and landscape heterogeneity scales and is well suited for application to eddy covariance measurements since all necessary information can be modelled or obtained from a few additional measurements.

## 1 Introduction

Understanding how energy in the form of sensible and latent heat is exchanged between the biosphere and the atmosphere is of great importance in different fields. For example, it is critical for weather forecasting and climate modelling [1–4], understanding of $CO_2$ sequestration by plants [5–9], and developing management recommendations for pastures, croplands, and forestry that enable efficient use of water resources [10–15]. Traditionally, the eddy covariance (EC) method is the approach used to measure the momentum, energy, and mass fluxes between the earth's surface and the atmosphere [16,17]. This is the most direct and non-destructive method to quantify momentum, energy, and mass fluxes between a given ecosystem and the atmospheric boundary layer (ABL) [18]. Nonetheless, it has been repeatedly found that when experimentally measuring the surface energy balance (SEB, balance between the energy reaching/leaving the ground surface through net radiation and the corresponding ground and turbulent fluxes, storage, and metabolic terms) there is a 10–30 % imbalance, resulting in an energy gap in the SEB [19–21].

Multiple possible causes for this gap have been investigated over the years. Some of them are instrument errors, including systematic error of sonic anemometers or humidity measurements [22–27], systematic error in measurements of other SEB components like soil heat flux or radiation [28–30], footprint mismatch [31], and heat storage in tall vegetation canopies [32–35]. These error sources have been progressively addressed by improving the measurement techniques and the development of correction methods that can be applied during data post-processing [36–38]. Nonetheless, besides all these significant efforts there remains an important SEB gap [37,39,40].

The EC method relies on very high temporal resolution measurements of the three-dimensional wind speed and any other additional scalar of interest. For example, if one is interested in measuring the sensible heat flux, then the temperature would be the additional scalar of interest. The sensible heat flux is then calculated as the covariance of the vertical wind speed and the temperature around the average for a defined time period of typically 30 minutes [17,41]. As a result, the EC method can only capture the turbulent contribution of the energy fluxes, where the turbulent fluctuations are defined around the adopted averaging period [42]. Initially, this was thought to be sufficient given that almost all atmospheric transport in the boundary layer is considered of turbulent nature [43,44]. However, more recently, it was found that, under certain atmospheric conditions, a significant part of the energy is not transported by turbulent motions but rather large-scale persistent atmospheric circulations that contribute to the vertical mean wind and reach far into the atmospheric surface layer [45,46]. This transport of sensible and latent heat by secondary circulations can significantly contribute to the SEB non-closure [30,40]. However, because it is expressed through a mean advective transport in the differential equations, it can only be captured through a spatial array of sensors and not by single-tower EC measurements [47–49]. At present, it is possible to differentiate between two types of secondary circulations. The first type are the so-called thermally-induced mesoscale circulations (TMCs) which result from heterogeneous surface forcing and are therefore spatially bound [30,40,50,51]. The second type are slow-moving turbulent organized structures (TOSs) that develop randomly even over homogeneous surfaces [52,53] and can translate with time. Recent data analysis has shown that extending the averaging period to several days instead of 30 minutes could almost close the energy balance gap for some sites [37,54] but not for all [41]. This could be explained because the TMCs are bound to the surface and thus do not move over time [40,53,55]. Moreover, such long averaging periods typically violate the stationarity requirement that has to be fulfilled to calculate a covariance [37,40].

Multiple approaches to correct for the SEB non-closure have been developed already, e.g. by extending the averaging period [37,41,54] applying the Bowen ratio of the measured turbulent fluxes to the missing dispersive fluxes [38], attributing the entire residual to the sensible [56] or latent [57] heat flux, or modelling the energy balance gap [58–60]. While some of these correction methods have proven to improve SEB closure [61–63], these models do not consider the factors and processes that cause the SEB gap. Some approaches consider the influence of atmospheric stability or heterogeneity in surface roughness [59,60] (Sect. 2), but they do not take into account the influence of thermal surface heterogeneity.

We hypothesize that it is possible to overcome the SEB non-closure problem by considering both, the influence of thermal landscape heterogeneity, and the effect of atmospheric stability. Our study expands beyond the earlier LES works of De Roo et al. [58], and Margairaz et al. [64]. Specifically, we use the correction method developed by De Roo et al. [58] that models the SEB non-closure as a function of the atmospheric stability factor $u_*/w_*$ (here $w_*$ is the Deardorff velocity) and take it one step further by including the effect of landscape heterogeneity. For this second step, we use the thermal heterogeneity parameter defined in Margairaz et al. [64]. The use of the LES technology is ideally suited to investigate the influence of atmospheric stability and surface heterogeneity on the SEB gap because it allows the control of both, the atmospheric conditions, and the surface characteristics. This facilitates the development of idealized analysis that can later shed light in the datasets of more complex field experiments [53,65,66]. Furthermore, LESs provide information on the structure of the atmospheric flow as a function of time, and the contribution of turbulent and advective transport of latent and sensible heat fluxes at each point in space [53,65].

This paper is organized as follows. In section 2, we provide a brief overview of former LES-based energy balance closure approaches, the two studies by De Roo et al. [58] (Sect. 2.1) and Margairaz et al. [64] (Sect. 2.2), and the theory underlying our new model (Sect. 2.3). Section 3 describes the LES. In Section 4, we present the resulting reference models (Sect. 4.1) and our new model (Sect. 4.2), which are further discussed in Section 5. Section 6 provides a short summary of our findings.

## 2 Theory

Several field studies have investigated EC measurements at multiple sites to understand the systematic behavior of the SEB closure, and have found relations with surface inhomogeneity [60,67–70], friction velocity $u_*$ [19,20,71,72], and atmospheric stability [19,20,72,73]. Also, large-eddy simulation (LES) studies confirm the dependence of the SEB gap with surface heterogeneity [74], $u_*$ [65] and atmospheric stability [59,75]. The relation between the SEB gap and surface heterogeneity can be explained as follows: the patches in heterogeneous surfaces heat up differently, which favors the formation of TMCs in addition to TOSs, with the amplitude and size of the individual surfaces conditioning how strongly these TMCs will be [53,66,76,77]. There is also a causal relation between the SEB gap and atmospheric stability: a large horizontal geostrophic wind speed, i.e., neutral to stable atmospheric stratification, results in enhanced horizontal mixing, which is why the influence of TOSs and TMCs on the measured flux is less pronounced than under free convective conditions [65,78].

At present, there exists only a reduced set of approaches to model the SEB closure based on the underlying processes by considering the factors that determine the magnitude of the energy balance gap such as atmospheric stability or surface heterogeneity. One of them is the model of Huang et al. [59] that depends on $u_*$ and $w_*$, the measurement height $z$, and the atmospheric boundary layer height $z_i$. This model is applicable to 30-min flux measurements, but it was only developed for homogeneous surfaces, and only larger heights were considered, so it is not applicable to typical EC measurement heights within the surface layer [58].

Another model is the one of Panin and Bernhofer [60]. They developed a heterogeneity-dependent energy balance gap parametrization that depends on changes in surface roughness and a corresponding heterogeneity length scale. However, this model does not include the effect of thermal heterogeneity [45,53]. Furthermore, it does not account for the effect of changing atmospheric conditions [30,50] and only provides the average energy balance closure for a site. As a result, it is rarely applicable to 30-min flux measurements.

**2.1 The atmospheric stability dependent energy balance gap model of De Roo et al.**

De Roo et al. [58] developed a parametrization for the SEB gap within the surface layer that results from the energy transport by TOSs. The so-called imbalance ($I$) is a suitable measure of the missing part of the energy fluxes, i.e., the advective and dispersive fluxes that do not contribute to the Reynolds flux [58,59]. It is based on the flux balance ratio that is computed as the Reynolds flux $H$ divided by the total available flux, which equals the surface flux $H_s$ in LES, and defined as

$$I = 1 - \frac{H}{H_s}. \tag{1}$$

Following the findings of Huang et al. [59], De Roo et al. [58] assumed that the underestimation of the heat fluxes, i.e. the imbalance $I$ can be described by a function of the non-dimensional scaling parameter $u_*/w_*$, as well as a function of the measurement height $z$ relative to the boundary layer height $z_i$:

$$I = F_1\left(\frac{u_*}{w_*}\right) F_2\left(\frac{z}{z_i}\right). \tag{2}$$

To determine the shape of functions $F_1$ and $F_2$, they developed an LES dataset of ABL flow over idealized homogeneous surfaces using PALM [79]. They considered nine combinations of atmospheric stability and Bowen ratios (Bo) with a vertical grid spacing of only 2 m to investigate the energy imbalance at a height of 0.04 $z_i$, i.e. within the atmospheric surface layer where most EC stations around the world are employed [19,80].

They found that combining two sets of scaling functions described well the imbalances in the sensible and latent heat fluxes. Specifically, they found that the sensible heat flux imbalance within the surface layer can be described with

$$F_{1,DR} = 0.197 \exp\left(-17.0 \frac{u_*}{w_*}\right) + 0.156 \tag{3}$$

and

$$F_{2,DR} = 10.69 \frac{z}{z_i} + 0.21. \tag{4}$$

**2.2 The thermal heterogeneity parameter of Margairaz et al.**

As part of the Idealized Planar Array study for Quantifying Spatial heterogeneity (IPAQS) [70], Margairaz et al. [81] developed a set of idealized LES of convective boundary layers over homogenously rough surfaces with embedded thermal heterogeneities of different scales. In their work, a wide range of mean geostrophic wind was implemented to vary the flow characteristics from inertia driven to buoyancy dominated. The goal of the study was to determine under what flow conditions TMCs are formed and to unravel the relation between the surface heterogeneity length scales and the dynamic length scales characterizing the TMCs. In their work, the authors show how TMCs express through mean advective transport of heat, which when unresolved either due to coarse numerical grid resolution or coarse experimental distribution of sensors can then be equivalently expressed through dispersive fluxes [81].

Furthermore, in their work, a scaling analysis between the vertical mean momentum equation and the continuity equation lead the authors to a non-dimensional parameter, referred to therein as heterogeneity parameter, that was shown to scale well with the contribution of dispersive fluxes when normalized by the turbulent fluxes [64]. The thermal heterogeneity parameter developed therein not only depends on the horizontal heterogeneity length scale $L_h$ but also on the length scale characteristic of the TMCs, $L_d$ which also depends on buoyancy and the mean horizontal wind speed. Specifically, the thermal heterogeneity parameter was defined as

$$\mathcal{H} = \frac{g\, L_h}{U_g^2} \frac{\Delta T}{\langle \overline{T_s} \rangle}, \tag{5}$$

where $T_s$ is the surface temperature and $\Delta T$ is the amplitude of the surface temperature heterogeneities, calculated following

$$\Delta T = \langle |\overline{T_s} - \langle \overline{T_s} \rangle| \rangle. \tag{6}$$

Interestingly, the heterogeneity parameter can also be interpreted as a modified Richardson number, representing a balance between the mean buoyancy forces developed by the thermal surface heterogeneities, and the inertia forces represented by the geostrophic wind that tend to blend the surface effects.

In this work, we will revisit the scaling relation from De Roo et al. [58] developed for homogeneous surfaces, and the one from Margairaz et al. [64] for heterogeneous surfaces, and we will illustrate how they complement each other and can be generalized to a single relation valid for both, TMCs and TOSs. Results from the work presented herein will therefore lead to a generalization of the correction scaling relation for the closure of the SEB presented initially in De Roo et al. [58].

**2.3 The combination of the atmospheric stability and thermal heterogeneity parameters into a new model**

To our knowledge, no existing approach considers both the influence of atmospheric stability and thermal surface heterogeneity on the magnitude of the SEB gap. The SEB model based on the atmospheric stability of De Roo et al. [58] and the thermal heterogeneity parameter developed by Margairaz et al. [64] proved to capture the changes in the magnitude of the SEB very well. We hypothesize, that combining their findings in one model will lead to a very powerful tool to parameterize the SEB gap in EC measurements. This new model could then be applied to various combinations of atmospheric stability and surface heterogeneity found in numerous eddy covariance measurements worldwide.

In this work, the focus is placed on the atmospheric surface layer (ASL) because eddy-covariance measurements are typically carried out close to the ground, within the surface layer [19,80]. Correspondingly, the analysis is carried out at the height of $z = 0.04\ z_i$, which corresponds to 52–59 m above the surface in our simulations. We calculate the imbalance ratio as defined in De Roo et al. [58] following Eq. 1. Specifically, the turbulent flux, $H$, is calculated using the 30-min averaged values of vertical wind speed $w$ and temperature $\theta$, as well as the 30-min averaged temporal covariance of $w$ and $\theta$ and the subgrid-scale contribution $H_{sgs}$,

$$H = \langle \overline{w\theta} - \overline{w}\,\overline{\theta} + \overline{H_{sgs}} \rangle. \tag{7}$$

The overbars indicate temporal averaging and the angled brackets denote horizontal averaging over the entire extent of the domain. In contrast to De Roo et al. [58], we therefore use the horizontally averaged imbalance instead of the local one. The sensible surface heat flux at the ground $H_s$ corresponds to $H$ at the lowest grid point ($dz/2$).

To parametrize the imbalance, we first produce a set of reference models by adapting the existing model of De Roo et al. [58] to each heterogeneity scale in our dataset as described in Sect. 2.3.1. Then, we proceed with developing the new model by including another scaling function that accounts for the influence of heterogeneity (Sect. 2.3.2).

**2.3.1 Parametrization of the Imbalance with respect to atmospheric stability (reference models)**

First, we adapt the existing model of De Roo et al. [58] to each of the datasets to obtain a benchmark for our new model. This results in four $F_1$ scaling functions that are similar to the scaling function presented in De Roo et al. [58], but represent one heterogeneity case, respectively. Following De Roo et al. [58], we factorize the imbalance following Eq. 1, assuming that the imbalance can be described by two scaling functions that are functions of the stability parameter $u_*/w_*$ and the normalized measurement height $z/z_i$. Based on the findings by De Roo et al. [58], we first assume that $F_1$ is an exponential function of the form $F_1 = a\ exp(b\ u_*/w_*) + c$, and $F_2$ is a linear function of the form $F_2 = i\ z/z_i + j$, where $a, b, c, i, j$, are fitting constants. Thus, we first fit $F_1$ to each of the simulation sets, individually, and later, we fit all of them onto a single $F_2$ function to observe their collapse on a unique curve.

For this analysis, we calculate the friction velocity $u_*$ and the Deardorff velocity $w_*$ directly using the 30-min averaged covariances as it would be done with experimental data obtained from EC systems. Thus, we calculate $u_*$ following

$$u_* = \langle \left(\overline{u'w'}^2 + \overline{v'w'}^2\right)^{1/4} \rangle, \tag{8}$$

where $u$ and $v$ are the horizontal wind speeds in x- and y-direction, and $w_*$ following

$$w_* = \langle \left(\frac{g}{\bar{\theta}} z_i \overline{w'\theta'}\right)^{1/3} \rangle, \tag{9}$$

where $g$ is the gravitational acceleration (9.81 m s$^{-2}$). Here, $z_i$ is determined as the height at which the total sensible heat flux crosses the zero value prior to reaching the capping inversion.

The resulting set of four $F_1$ scaling functions for each of the datasets and one $F_2$ scaling function for all of the datasets is then used as a benchmark for our new model and referred hereafter as reference models.

**2.3.2 Parametrization of the Imbalance with respect to atmospheric stability and surface heterogeneity (new model)**

To consider the effect of surface heterogeneity, we assume that instead of describing the imbalance with a different scaling function $F_1$ for each set of simulations, it is possible to use the scaling function that describes the imbalance in the simulations with a homogeneous surface, $F_{1,HM}$, and add another scaling function, $F_3$, that accounts for the heterogeneity:

$$I\left(z, \frac{u_*}{w_*}\right) = F_{1,HM}\left(\frac{u_*}{w_*}\right) F_2\left(\frac{z}{z_i}\right) F_3(\mathcal{H}), \tag{10}$$

where $\mathcal{H}$ is the thermal heterogeneity parameter introduced in Margairaz et al. [64] (Eq. 4).

After analyzing the relationship between $I$ normalized with $F_{1,HM}$, we assume that the relationship between $I/F_{1,HM}$ and $\mathcal{H}$ is of linear nature and fit $I/F_{1,HM}$ to $F_3 = m\mathcal{H} + n$. Once $F_3$ is found, we proceed to identify the new $F_2$ similarly to the previous section.

## 3 Dataset and study cases

The data used in this study was originally developed in the computational work of Margairaz et al. [64,81]. They used the pseudo-spectral LES approach that was first introduced by Moeng [82] and Albertson and Parlange [83] and further developed by Bou-Zeid et al. [84], Calaf et al. [85], and Margairaz et al. [86]. The data consists of a set of numerical simulations of a characteristic ABL developed over a homogeneously rough and flat surface. The simulations represent an idealized dry ABL, forced through a geostrophic wind at the top with Coriolis force, and an imposed surface temperature at the bottom of the domain.

Study cases include a set of simulations with homogenous surface temperature (referred hereafter as HM) and a second set with heterogeneous surface temperature distributions (referred hereafter as HT). In both sets, the geostrophic wind is varied between 1 m s$^{-1}$ to 15 m s$^{-1}$. For the set of heterogeneous surface temperature conditions, the corresponding length scale of the characteristic surface heterogeneities is also varied, considering cases with 800 m, 400 m, and 200 m patches (referred hereafter as HT200, HT400, HT800, see figure 1). In this case, the surface temperature variations are randomly distributed following a gaussian distribution with a standard deviation of ± 5 K and mean temperature equal to that of the homogenous cases, namely 290 K. In all cases the surface temperature is initialized at a temperature of 5 K higher than the air temperature to promote the development of a convective boundary layer. All simulations have a domain size $(l_x, l_y, l_z) = (2\pi, 2\pi, 2)$ km with a horizontal grid-spacing of $\Delta x = \Delta y = 24.5$ m and a vertical grid-spacing of $\Delta z = 7.8$ m, resulting in $(N_x, N_y, N_z) = (256, 256, 256)$ grid points. At the bottom boundary, the surface heat flux is computed from the imposed surface temperature $\theta_s$ using Monin-Obukhov similarity theory.

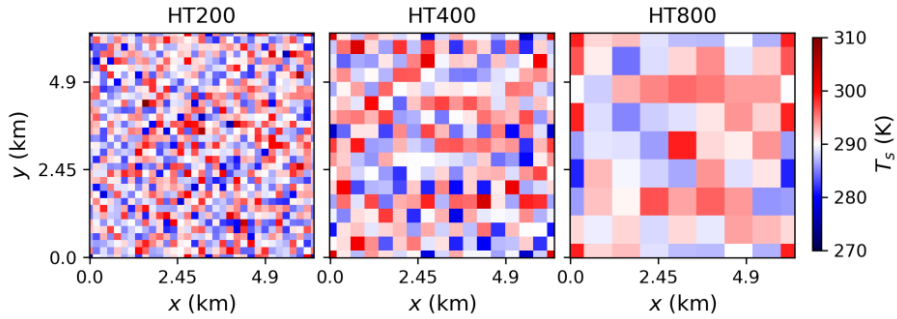

**Fig 1.** Temperature distribution at the surface for the three sets of simulations with heterogeneous surfaces.

In all cases, the initial boundary layer height $z_i$ was set to 1000 m by applying a capping inversion of 0.012 K m$^{-1}$. While $\theta_s$ remains stable over the entire simulation time, the air temperature increases over time, leading to slightly less unstable atmospheric conditions over time. However, this effect was found to be negligible over the short duration of the simulation [64].

In total, 28 simulations were performed with different atmospheric conditions, controlled by seven different geostrophic wind speeds (i.e. $U_g$ = 1, 2, 3, 4, 5, 6, 9, 15 m s$^{-1}$) for each set of homogeneous and heterogeneous surface conditions. In the simulations, the Coriolis parameter was set to 10$^{-4}$ Hz, representative of a latitude of 43.3° N. Also, the roughness length was set to 0.1 m for all simulations, and the used thermal roughness was set to 1/10 $z_0$ following [87]. More details on the numerical simulations can be found in the original work of Margairaz et al. [64].

For the analysis presented in this work, we use statistics over a 30-minute interval recorded after 4 hours of spin-up time.

## 4 Results

### 4.1 Reference models

As described in section 2.3.1, we first fitted the exponential function $F_1$ to each one of the simulation cases resulting in four different sets of parameters, shown in Table 1 for the scaling function $F_{1,h}$:

$$F_{1,h} = a_h \exp\left(b_h \frac{u_*}{w_*}\right) + c_h. \tag{11}$$

Note that for each simulation case, there exist seven data points corresponding to the changes in geostrophic forcing and hence to different thermal stratification. These four different fits describe the imbalance ratio for each surface heterogeneity condition as a function of the non-dimensional term $u_*/w_*$.

Table 1. Fitting parameters for all simulation cases with different surface characteristics.

| h | $a_h$ | $b_h$ | $c_h$ |
|---|---|---|---|
| HM | 0.133 | -15.3 | 0.056 |
| HT200 | 0.203 | -12.6 | 0.058 |
| HT400 | 0.289 | -10.2 | 0.055 |
| HT800 | 0.435 | -14.3 | 0.073 |

The values calculated for $u_*/w_*$ are shown in Table 2 where all relevant parameters characterizing the simulations are summarized. Figure 2a shows that these fitted functions for $I$ collapse into the same value of roughly 6 % of $H_s$ under less unstable conditions ($u_*/w_* > 0.4$). Only for HT800, the imbalance ($I$) settles at around 8 % of $H_s$ for the weaker unstable conditions. Alternatively, the imbalance increases with increasing instability, with the weakest increase found in the homogeneous surface cases and stronger increases with heterogeneous surfaces. The increase also depends on the patch size, being strongest with the largest patch size.

Table 2. Overview over characteristic variables that are relevant for this study, including geostrophic wind speed $U_g$, boundary layer height $z_i$, the friction velocity $u_*$, the Deardorff velocity $w_*$, the atmospheric stability parameters $u_*/w_*$ and $-z_i/L$, the heterogeneity parameter $\mathcal{H}$, and the energy imbalance $I$ for each simulation.

| Name | $U_g$ (m s$^{-1}$) | $z_i$ (m) | $u_*$ | $w_*$ | $u_*/w_*$ | $-z_i/L$ | $\mathcal{H}$ | $I$ (%) |
|---|---|---|---|---|---|---|---|---|
| HM-1 | 1 | 1328 | 0.15 | 1.59 | 0.09 | 493.65 | 0 | 8.79 |
| HM-2 | 2 | 1318 | 0.22 | 1.57 | 0.14 | 156.17 | 0 | 7.28 |
| HM-3 | 3 | 1306 | 0.27 | 1.56 | 0.17 | 75.30 | 0 | 6.40 |
| HM-4 | 4 | 1293 | 0.33 | 1.54 | 0.21 | 42.30 | 0 | 6.02 |
| HM-6 | 6 | 1295 | 0.43 | 1.55 | 0.28 | 18.58 | 0 | 6.08 |
| HM-9 | 9 | 1307 | 0.57 | 1.59 | 0.36 | 8.63 | 0 | 5.60 |
| HM-15 | 15 | 1340 | 0.81 | 1.64 | 0.49 | 3.33 | 0 | 5.46 |
| HT200-1 | 1 | 1439 | 0.11 | 1.55 | 0.07 | 1230.97 | 27.09 | 14.38 |
| HT200-2 | 2 | 1434 | 0.17 | 1.58 | 0.11 | 328.31 | 6.77 | 10.86 |
| HT200-3 | 3 | 1430 | 0.24 | 1.63 | 0.15 | 128.00 | 3.01 | 9.22 |
| HT200-4 | 4 | 1434 | 0.30 | 1.67 | 0.18 | 67.68 | 1.69 | 8.01 |
| HT200-6 | 6 | 1402 | 0.41 | 1.72 | 0.24 | 28.75 | 0.75 | 6.49 |
| HT200-9 | 9 | 1374 | 0.56 | 1.79 | 0.31 | 12.88 | 0.33 | 6.31 |
| HT200-15 | 15 | 1383 | 0.81 | 1.96 | 0.41 | 5.60 | 0.12 | 5.92 |
| HT400-1 | 1 | 1487 | 0.09 | 1.55 | 0.06 | 1951.94 | 52.14 | 21.14 |
| HT400-2 | 2 | 1463 | 0.17 | 1.63 | 0.11 | 336.34 | 13.03 | 15.34 |
| HT400-3 | 3 | 1452 | 0.24 | 1.65 | 0.14 | 135.45 | 5.79 | 13.28 |
| HT400-4 | 4 | 1434 | 0.30 | 1.68 | 0.18 | 67.56 | 3.26 | 9.56 |
| HT400-6 | 6 | 1406 | 0.41 | 1.72 | 0.24 | 29.07 | 1.45 | 7.09 |
| HT400-9 | 9 | 1390 | 0.56 | 1.80 | 0.31 | 13.02 | 0.64 | 6.59 |
| HT400-15 | 15 | 1404 | 0.81 | 1.95 | 0.42 | 5.59 | 0.23 | 6.52 |
| HT800-1 | 1 | 1487 | 0.10 | 1.57 | 0.07 | 1428.40 | 79.16 | 25.07 |
| HT800-2 | 2 | 1459 | 0.17 | 1.59 | 0.11 | 338.84 | 19.77 | 14.66 |
| HT800-3 | 3 | 1448 | 0.24 | 1.62 | 0.15 | 117.37 | 8.8 | 12.59 |
| HT800-4 | 4 | 1431 | 0.30 | 1.63 | 0.19 | 60.98 | 4.95 | 12.49 |
| HT800-6 | 6 | 1397 | 0.41 | 1.65 | 0.25 | 26.56 | 2.2 | 8.21 |
| HT800-9 | 9 | 1394 | 0.56 | 1.69 | 0.33 | 10.93 | 0.98 | 7.20 |
| HT800-15 | 15 | 1407 | 0.81 | 1.79 | 0.46 | 4.22 | 0.35 | 6.83 |

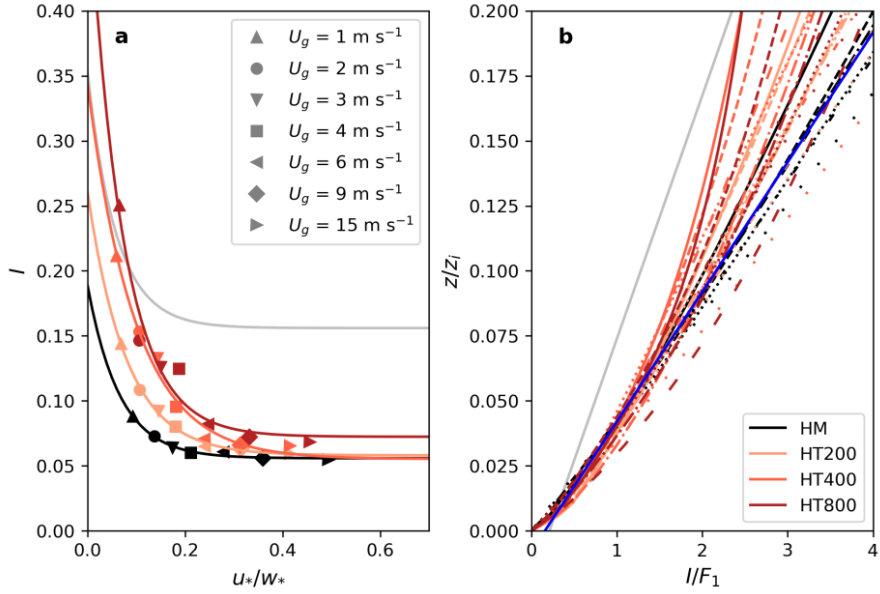

**Fig 2.** Panel a shows the imbalance ratio $I$ at 0.04 $z/z_i$ as a function of the stability parameter $u_*/w_*$. The four simulation sets are represented by different colors. For each simulation set, a separate fit of the scaling function $F_1$ was performed, resulting in Eq. 11 with different fitting parameters shown in Table 1. The atmospheric stability is steered by changes in $U_g$, shown by different marker shapes. The grey line shows the fit obtained by De Roo et al. [58] (Eq. 3) for comparison. Panel b shows the vertical profiles of the imbalance normalized with the four fits of $F_1$ (Eq. 11), respectively. The blue line shows the fitted scaling function $F_{2,R}$ (Eq. 12). Again, the respective scaling function derived by De Roo et al. [58] (Eq. 4) is shown in grey for comparison.

We then normalized the imbalance ratios (i.e. Eq. 1, also vertical axis in Fig. 2a) with the four different scaling functions for the respective simulations (Eq. 11, Table 1). Results are then represented in Fig. 2b as a function of the second non-dimensional term identified in De Roo et al. [58], namely $z/z_i$. At this stage, the data presents a nice unique collapse for $z/z_i < 0.1$, representative of the surface layer region.

This unified scaling is well represented by function $F_{2,R}$

$$F_{2,R} = i_R \frac{z}{z_i} + j_R, \tag{12}$$

where $i_R$ is 20.05 and $j_R$ is 0.157.

## 4.2 New model

Figure 3 also shows the normalized imbalances, but in this case, the scaling function that was derived for the homogeneous simulations ($F_{1,HM}$, Eq. 11, Table 2) was used for all simulations. Here, the profiles don't collapse into a single curve, but instead present a data spread, with the largest deviation found once again in the $L_h = 800$ m configuration.

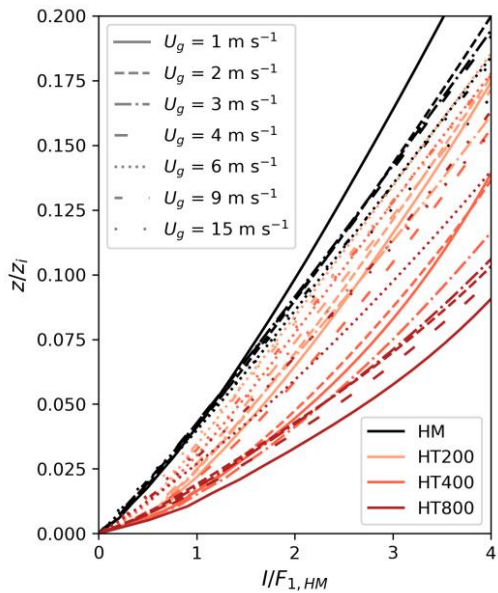

**Fig 3.** Vertical profiles of the imbalance all normalized with the same scaling function $F_{1,HM}$ (Eq. 11, Table 2).

Next, we investigate whether these deviations can be reduced if the imbalance ($I$) is normalized by $F_{1,HM}$ and represented as a function of the thermal heterogeneity parameter ($\mathcal{H}$). In this case, figure 4 shows that two different linear relationships can be differentiated for those cases with weak geostrophic forcing ($U_g = 1$ m s$^{-1}$) and those with a more moderate or stronger wind ($U_g \geq 3$ m s$^{-1}$). We find those two groups to correspond to the formation of cellular and roll-like secondary circulations.

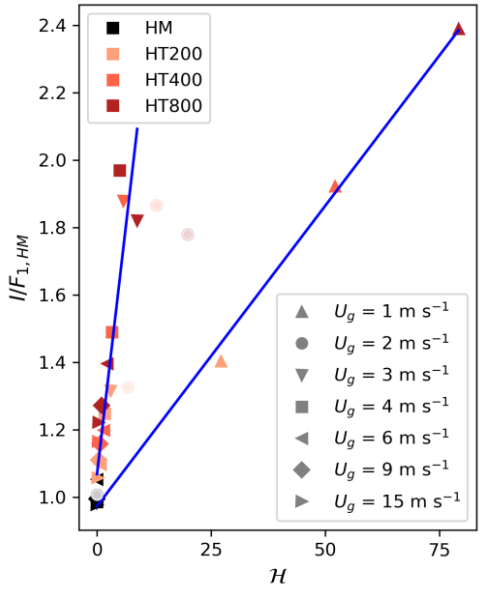

**Fig 4.** Dependence of the imbalance all normalized with the same scaling function $F_{1,HM}$ (Eq. 11, Table 1) on the heterogeneity parameter $\mathcal{H}$. The data is separated into two groups: (1) simulations with $U_g = 1$ m s$^{-1}$ that show cellular shaped secondary circulations and (2) simulations with $U_g \geq 3$ m s$^{-1}$ simulation that show roll-shaped secondary circulations. Simulations with $U_g = 2$ m s$^{-1}$ are discarded because they show no clearly cellular nor roll-shaped structures. The two blue lines show the fits of the third scaling function to the two groups ($F_{3,c}$ and $F_{3,r}$, Eq. 13–14)

Figure 5 shows xy-cross-sections of the 30-min averaged vertical wind speed $w$ for combinations of $U_g = 1, 2, 3, 4$ m s$^{-1}$ and $L_h = 200, 400$ m. While in the case of $U_g = 1$ m s$^{-1}$ (Fig. 5a,e) there are large cellular circulations taking place, they disappear with increasing wind speed to give place to the formation of roll-type turbulent structures for $U_g > 3$ m s$^{-1}$ (Fig. 5c-d,g-h). The structures resulting from $U_g = 2$ m s$^{-1}$ (Fig. 5b,f) are neither cellular nor roll-like and are therefore excluded from the analysis.

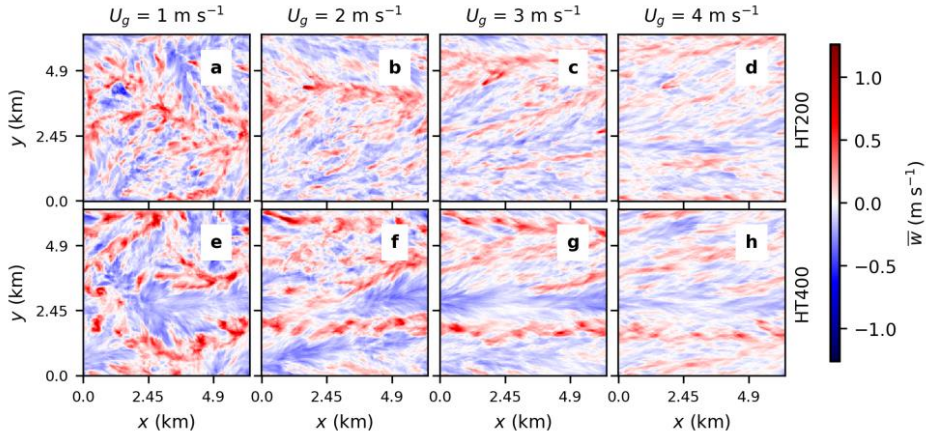

**Fig 5.** Exemplary horizontal cross-sections of the half-hourly averaged vertical wind speed $w$ at 0.04 $z_i$ for HT200 and HT400 simulations with $U_g$ up to 4 m s$^{-1}$.

Fitting $F_3$ to the two datasets, we receive the following scaling functions:

$$F_{3,c} = m_c \, \mathcal{H} + n_c, \tag{13}$$

with $m_c = 0.018$ and $n_c = 0.973$ for $u_*/w_* < 0.1$, which is valid for all simulations where cellular structures develop ($U_g = 1$ m s$^{-1}$), and

$$F_{3,r} = m_r \, \mathcal{H} + n_r, \tag{14}$$

with $m_r = 0.116$ and $n_r = 1.07$ for $u_*/w_* > 0.14$, which is valid for all simulations where roll-like structures develop ($U_g \geq 3$ m s$^{-1}$). The fit for the very unstable simulations ($u_*/w_* < 0.1$) describes the normalized imbalance with a very high $R^2$ of 0.996. For the corresponding fit to the less unstable conditions ($u_*/w_* > 0.14$), the $R^2$ value is slightly lower with 0.841.

When normalizing the imbalance additionally with $F_{3,c}$ or $F_{3,r}$, respectively, the vertical profiles of imbalance collapse similar to when they are normalized with different $F_1$ scaling functions for each stability, as shown in Figure 6. In this case, the remaining imbalance can be described following Eq. (15):

$$F_{2,N} = i_N \frac{z}{z_i} + j_N, \tag{15}$$

with $i_N = 20.2$ and $j_N = 0.153$.

All characteristic variables that are relevant for our simulations are summarized in Table 2.

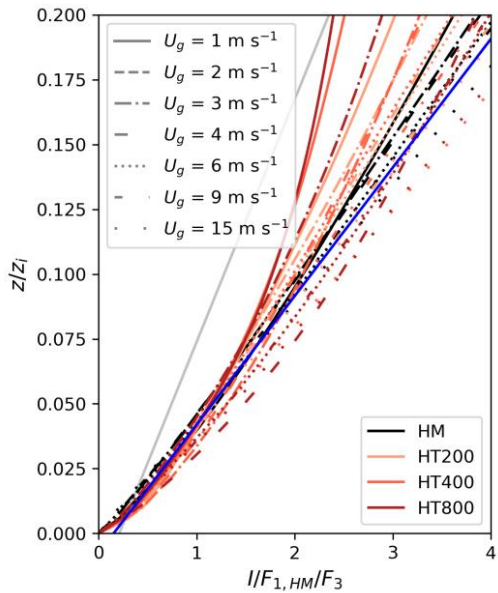

**Fig 6.** Vertical profiles of the imbalance all normalized with the same scaling function $F_{1,HM}$ (Eq. 11, Table 1) and the respective scaling functions $F_{3,c}$ or $F_{3,r}$ (Eq. 13–14). The blue line shows the fitted scaling function $F_{2,N}$ (Eq. 15). The scaling function derived by De Roo et al. [58] is shown in grey for comparison.

## 5 Discussion

The reference models derived by fitting one curve for each heterogeneity scale are a more direct way to parametrize the energy imbalance than the new model, as they rely on fewer assumptions. Specifically, they are tailored to each heterogeneity scale and do not rely on the additional assumption that the magnitude of the SEB gap relates to the heterogeneity scale. However, it is not practical to use as a correction method to real measurements because it is only applicable to the discrete heterogeneity scales covered in this study. This means, that for each study case, there is a need to re-derive the corresponding scaling relation.

Alternatively, using the new method proposed in this work that adds a third scaling function to parametrize the imbalance as a function of the heterogeneity parameter to account for the surface characteristics which facilitates the generalization of the correction method to EC towers surrounded by landscapes featuring any characteristic heterogeneity scale. However, because our dataset only covered heterogeneity length scales up to $L_h = 800$ m, which corresponds to 0.57 $z_i$ on average, it is questionable whether the resulting scaling function $F_3$ would hold for larger heterogeneity length scales. Zhou et al. [76] investigated the relation between the scale of surface heterogeneity and the SEB gap and found that the SEB gap increases with heterogeneity length scale, reaching its maximum when the heterogeneity length scale is of the order of the boundary layer height, and decreases again, with even larger heterogeneity length scales. Our results confirm that the imbalance increases with the heterogeneity scale, especially under very unstable conditions (Fig. 2a, Table 2) but based on the findings by Zhou et al. [76], we recommend using this scaling function only for $L_h \leq z_i$.

While the new model is very flexible regarding the landscape heterogeneity scale, it is not applicable to all atmospheric conditions. This is because we were unable to define the scaling function $F_3$ for the atmospheric conditions that cause sub-mesoscale circulations to form neither uniquely cellular nor roll-shaped. While Margairaz et al. [81] found that different geostrophic forcing leads to clearly cellular or roll-like structures using the roll factor defined by Salesky et al. [88], there is a transition zone in which the structures could not be clearly assigned to a cell or roll regime. In our analysis, we therefore excluded the simulations with $U_g = 2$ m s$^{-1}$. Several studies found the transition from cellular to roll-like structures to be rather sharp, occurring somewhere between $-z_i/L = 4.5$ and $-z_i/L = 45$ [89] or $-z_i/L = 8$ and $-z_i/L = 65$ [90], or at around $-z_i/L = 25$. Other studies have found the transition to occur more gradually with transitional structures or co-existing rolls and cells for $-z_i/L = 14.1$ [91], or for $-z_i/L < 21$ [92]. For better comparison, we converted $u_*/w_*$ to $-z_i/L$ for our simulations using

$$-\frac{z_i}{L} = \kappa \left(\frac{u_*}{w_*}\right)^{-3}, \tag{16}$$

where $\kappa$ is the von Kármán constant (0.4) [93]. The resulting $-z_i/L$ values are shown in Table 2. For the simulations with $U_g = 2$ m s$^{-1}$, $-z_i/L$ varies between 156.17 and 338.84, indicating that the transition to clearly roll-like structures occurs at larger $-z_i/L$ values than reported by other studies.

To apply the correction method, a certain amount of information on the atmospheric conditions and the surrounding landscape is required. The atmospheric conditions are considered in $F_1$, using $u_*/w_*$ which can be calculated from the EC measurements similar to Eq. 8–9. $F_2$ is a function of $z/z_i$ which means that $z_i$ needs to be known, which cannot be derived from EC measurements, only. Mauder et al. [61] already tested the correction method proposed by De Roo et al. [58] using ceilometer measurements of $z_i$. For one site where no ceilometer measurements were available, they followed the method of Batchvarova and Gryning [94] to calculate $z_i$ using radiosonde measurements of the morning temperature gradient. They found the correction method leading to a good energy balance closure, even though the radiosonde measurements were taken at a distance of 170 km. Finally, the characteristic heterogeneity parameter can be derived using remote sensing methods or already available land cover maps [60]. At permanent measurement sites with continuous flux measurements, the temperature amplitude can be derived by performing ground-based measurements of surface temperature over the different landcover types surrounding the tower. In extensive measurement campaigns, additional airborne measurements can provide information on the temperature amplitude [80]. If additional measurements are too costly, however, it is also possible to model the surface temperature using the radiation measurements and landcover characteristics [95–97]. What is clear from these results, is that for accurate SEB studies, the use of single point measurements is not sufficient, but obtaining spatial information of the surroundings as well as from the flow is proven to be critical. This is a strong motivation for a paradigm change in the standard single point EC measurement approaches.

To compare the performance of our newly developed model with the reference models and with the parametrization developed by De Roo et al. 2018, we computed the imbalance using the scaling functions derived by the different approaches with

$$I = \frac{F_1 F_2 F_3}{1 - F_1 F_2 F_3} H. \tag{17}$$

The share of $H$ and $I$ in $H_s$, i.e. the total available heat flux, is shown in Fig. 7. Without any correction, $H$ is on average 90.24 ± 4.77 % of the total heat flux at 0.04 $z/z_i$. With the imbalance calculated using the reference models based on De Roo et al. [58], we obtain $H + I_R = 99.49 \pm 0.86$ %. Reaching nearly 100 % means that the energy balance gap is almost closed. At the same time, the standard deviation becomes significantly smaller, indicating that the method captured the deviations in the energy balance gap well. The use of our newly developed model for imbalance calculation gives similar results with $H + I_N = 99.53 \pm 0.87$ %. This shows that the newly developed model, which is much more flexible in its application to measurements, achieves just as good results as the reference models.

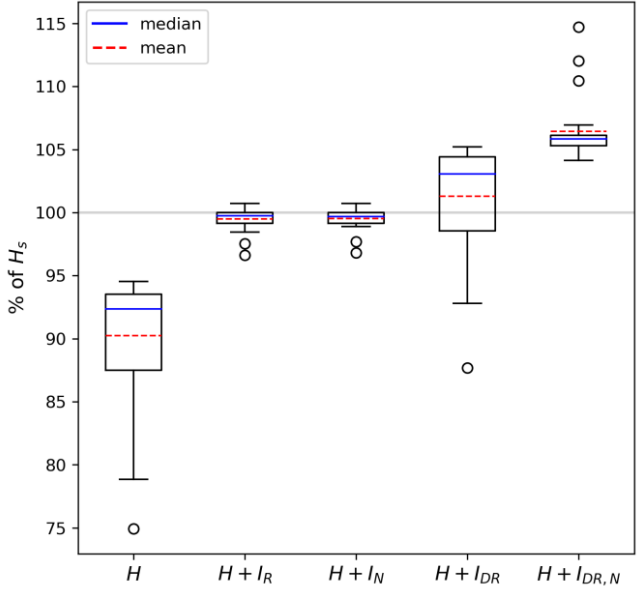

**Fig 7.** Comparison of different correction methods. The distribution of the heat flux H among all simulations is shown in box plots, where the blue line represents the median and the dashed red line represents the mean. At the very left, the distribution of the uncorrected H at 0.04 $z_i$ is shown. In second and third place, the H corrected with our approaches A and B are shown. In the fourth place, H corrected with the method by De Roo et al. [58] is shown. At the very right, H is corrected with a combination of $F_1$ and $F_2$ derived by De Roo et al. [58] and $F_3$ derived in approach B in this study.

Using the scaling functions defined by De Roo et al. [58] (Eq. 3–4) results in $H + I_{DR} = 101.28 \pm 4.2$ %. This shows that the method of De Roo et al. [58] generally works well with our data set, but it slightly overestimates the energy balance gap on average. This correction method has already been tested on EC measurements by Mauder et al. [61] who also found the method to yield good results. Furthermore, it does not capture the deviation of the imbalance due to heterogeneity as shown in Fig. 7, which is also reflected in the almost unchanged standard deviation. This is to be expected since this method was developed for homogeneous surfaces only. However, we do not recommend combining the scaling functions defined in De Roo et al. [58] and $F_3$ derived in this study to address the effect of the heterogeneity as it leads to a clear overcorrection with $H + I_{DR,N} = 106.34 \pm 2.41$ %.

## 6 Conclusion

We extended the energy balance gap correction method initially developed by De Roo et al. [58] taking into account the effects of spatial surface heterogeneity onto the atmospheric flow. We compared our new model to the reference models that are based on the already existing approach. The use of the reference models resulted in sets of two scaling functions for different heterogeneity scales, respectively. This approach is the more direct way to determine the imbalance and produces very good results. However, those sets of scaling functions are restricted to the distinct heterogeneity scales used in this study, which is why this approach is not transferable to all characteristic continuously distributed heterogeneity scales of the landscape surrounding an EC system, i.e. an area of about 20 × 20 km [20,60]. Our new model proved to yield similar results and its application to real-world EC tower sites is very flexible, since a third scaling function characterizing the influence of heterogeneity was introduced. Therefore, this correction method can be used for a wide range of characteristic heterogeneity scales of a landscape surrounding an EC tower. To apply the correction method, the atmospheric stability parameter $u_*/w_*$, the boundary layer height $z_i$, the heterogeneity scale $L_h$, and the amplitude of the surface temperature $\Delta T$ need to be known that can be either calculated from the EC measurements together with nearby operational radiosonde measurements or by using a ceilometer, and remotely-sensed land-surface-temperature data products.

**Acknowledgments**

This study was financially supported by Deutsche Forschungsgemeinschaft (DFG)Award #406980118 and the MICMoR Research School of KIT. Marc Calaf thanks the support of the National Science Foundation Grants PDM-1649067, and PDM-1712538 as well as the support of the Alexander von Humboldt Stiftung/Foundation, Humboldt Research Fellowship for Experienced Researchers, during the sabbatical year at the Karlsruhe Institute of Technology Campus Alpin in Garmisch-Partenkrichen.